\documentclass[aps,prl,twocolumn,groupedaddress, showpacs]{revtex4-1}
\usepackage{graphicx}
\usepackage{color}
\usepackage{amsmath}

\begin{document}


\title{Reciprocity Between Robustness of Period and Plasticity of Phase in Biological Clocks}


\author{Tetsuhiro S. Hatakeyama}
\email[]{hatakeyama@complex.c.u-tokyo.ac.jp}
\author{Kunihiko Kaneko}
\affiliation{Department of Basic Science, The University of Tokyo, \\
3-8-1 Komaba, Meguro-ku, Tokyo 153-8902, Japan}



\begin{abstract}
Circadian clocks exhibit the robustness of period and plasticity of phase against environmental changes such as temperature and nutrient conditions.
Thus far, however, it is unclear how both are simultaneously achieved.
By investigating distinct models of circadian clocks, we demonstrate reciprocity between robustness and plasticity: higher robustness in the period implies higher plasticity in the phase, where changes in period and in phase follow a linear relationship with a negative coefficient.
The robustness of period is achieved by the adaptation on the limit cycle via a concentration change of a buffer molecule, whose temporal change leads to a phase shift following a shift of the limit-cycle orbit in phase space.
Generality of reciprocity in clocks with the adaptation mechanism is confirmed with theoretical analysis of simple models, while biological significance is discussed.
\end{abstract}


\maketitle


Biological systems are both robust to external changes in the environment, and plastic to adapt to environmental conditions.
How are the robustness and plasticity, which seem to be opposing properties at a first glance, compatible with each other?
In the present Letter, we address this question, by focusing on biological clocks, which are ubiquitous in organisms.

Such biological clocks often work as pacemakers, to adapt to periodic events.
One of the most prominent examples of such oscillators is a circadian clock \cite{Dunlap1999, Bell-Pedersen2005}.
To respond to periodic events, the following two criteria are generally imposed on a biochemical oscillator.
\\
{\bf 1. Robustness of period:}
If the period of an oscillator strongly depends on external conditions, the oscillator would not accurately predict time.
For example, if the period of a circadian clock is sensitive to temperature, the clock malfunctions depending on the temperature.
To avoid such error, the period of pacemakers should not be affected by external conditions such as temperature and nutrient compensation \cite{Pittendrigh1954, Hastings1957}.
\\
{\bf 2. Plasticity of phase:}
The period of the circadian clock of most organisms is known not to correspond precisely with 24 hours \cite{Pittendrigh1976a}, and biological clocks are entrained with the external 24-hr cycle \cite{Pittendrigh1974}, so that the phase difference between the two does not increase with time.
This entrainment is also necessary to adapt an abrupt change in the environment that may cause temporal misalignment between the internal and external cycles.
For such entrainment, plasticity of the phase of the internal clock against external stimuli, e.g., changes in temperature and/or brightness, is needed.

Indeed, biological clocks satisfy both robustness and plasticity to changes in factors such as temperature and nutrient conditions, which change in the daily cycle.
For example, circadian clocks of {\it in vivo Drosophila} \cite{Zimmerman1968}, {\it Neurospora} \cite{Lakin-Thomas1990}, and  {\it in vitro} cyanobacteria \cite{Nakajima2005, Yoshida2009} show temperature compensation of a period and are entrained by cyclic temperature changes.
Robustness of period is also important to stable entrainment since it can reduce the difference between the period of inner clock and external cycle.
In spite of some studies discussing the compatibility between the two properties \cite{Zimmerman1968, Rand2006, Takeuchi2007, Akman2010}, however, little is known about the quantitative relationship between the two properties.

To answer how the robustness of the period and plasticity of phase are compatible with each other, we first analyze two major models of a circadian clock, i.e., post-translational oscillator (PTO) \cite{Tomita2005, Nakajima2005, Qin2010} and transcription-translation-based oscillator (TTO) \cite{Takeuchi2007, Akman2010, Qin2010}, which consists only of protein-protein interactions and both transcription and translation processes, respectively.
Without imposing any special mechanism, we demonstrate that biological clocks with robustness of period against changes in an environmental factor generally exhibit phase entrainment against the cyclic change of that factor --- reciprocity between the robustness of period and plasticity of phase: the plasticity increases with robustness.

For PTO model, we adopt the KaiC allosteric model \cite{VanZon2007}, for {\it in vitro} cyanobacterial circadian clock system \cite{Nakajima2005}.
Here, KaiC protein consists of six monomers, each of which has a phosphorylation site.
The protein has active and inactive forms.
Active (inactive) KaiC are phosphorylated (dephosphorylated) step by step, respectively.
Phosphorylation reactions are facilitated with KaiA as an enzyme and dephosphorylation reactions spontaneously progress without an enzyme.
$k_p$ and $k_{dp}$ denote the rate of phosphorylation and dephosphorylation of KaiC, respectively, which depend on temperature as $k_p \propto \exp(- \beta E_p)$ and $k_{dp} \propto \exp(- \beta E_{dp})$, where $E_p$ ($E_{dp}$) is the activation energy for phosphorylation (dephosphorylatiion), respectively, and $\beta$ is the inverse temperature by taking the Boltzmann constant as unity.
The temporal evolution of the concentration of each phosphorylated active (inactive) KaiC is given by rate equations (see model equations and Fig.1A of \cite{Supply}).

\begin{figure}[]
\includegraphics{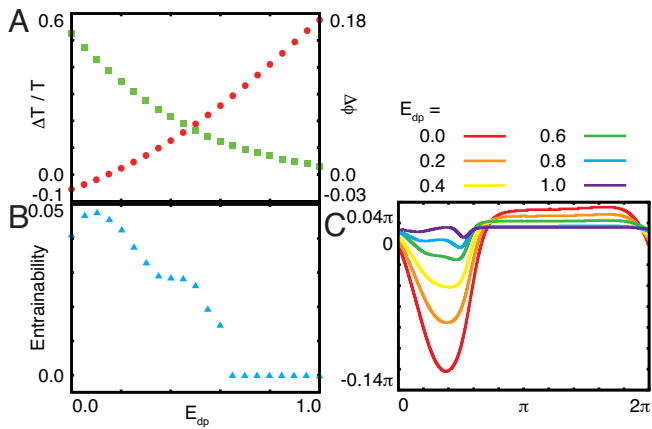}
\caption{
Reciprocity between the robustness of period and plasticity of phase in the PTO model.
(A) Difference between periods at two temperatures ($\beta_1 = 1.0$ and $\beta_2 = 1.5$) ($\Delta T / T$, red circle) and the amplitude of the phase response curve against a transient jump of temperature from $\beta_1$ to $\beta_2$ ($\Delta \phi$, green square) plotted against different values of $E_{dp}$ while $E_p$ is fixed at $1.0$.
$\Delta T$ is normalized by the period at $\beta_1$, and $\Delta \phi$ is normalized by the duration of stimulus and difference between $\beta_1$ and $\beta_2$.
$\Delta T / T$ and $\Delta \phi$ are negatively correlated across the entire range of $E_{dp}$.
(B)  Entrainability is plotted against various $E_{dp}$ with fixed $E_p = 1.0$.
}
(C)  Phase response curve against transient increase in $\beta$.
As a stimulus, the inverse temperature $\beta$ is increased from $\beta_1$ to $\beta_2$ for the duration of one unit of time.
Lines of different colors represent the PRCs for different values of activation energy for the dephosphorylation reaction $E_{dp}$.
\label{fig1}
\end{figure}

This model shows a limit-cycle attractor in which the total phosphorylation level, i.e., the ratio of phosphorylated monomers, oscillates in time.
We demonstrated that the robustness of the period against various environmental changes is achieved by enzyme-limited competition \cite{Hatakeyama2012, Hatakeyama2014}:
With the increase in temperature, the abundance of the active form of the KaiC molecule increases, which in turn decreases the abundance of the free KaiA molecule, and thus the increase in the rate of phosphorylation $k_p$ is canceled out, when the total KaiA amount,  $A_{total}$, is sufficiently small.
This robustness in the period is achieved when $E_p$ is sufficiently larger than $E_{dp}$.
We use the difference in periods between two different temperature conditions ($\Delta T / T$) as an indicator of the robustness of period.
Its dependence upon $E_{dp}-E_p$ is given in Fig.\ref{fig1}A.

This clock, on the other hand, entrains against external periodic change, so that the phase of phosphorylation oscillator coincides with that of external cycle.
By imposing external periodic change in temperature, we computed how many number of cycles are needed for the clock to entrain with this external cycle, and defined {\sl entrainability} as the inverse of the number (see \cite{Supply}). 
Dependence of the entrainability and $\Delta T / T$ upon $E_{dp}$ with fixed $E_p$ is plotted in Fig.\ref{fig1}A (red circle) and B.
As $E_{dp}-E_p$ is smaller, $\Delta T / T$ becomes smaller and the entrainability is higher.
In other words, if the period of the clock is more robust against temperature change, it is entrained faster with the external temperature cycle, i.e., the phase has higher plasticity.

Although this demonstrated the correlation between period robustness and phase plasticity, the entrainability here is a complicated indicator for the latter, as it can depend on the form of external cycle. 
Hence, we introduce a more tractable indicator for the plasticity of phase, by using a phase response curve (PRC) \cite{Winfree1980}.
PRC is a function of phase and represents a phase shift introduced by a transient stimulus.
When a transient stimulus is added to an oscillatory system, the period of oscillation is temporally altered depending on the phase when the stimulus was added. The period finally returns to its original value.
In this time, the phase of the oscillator progresses (or is delayed) from the original phase because of the temporal shortening (or lengthening) of the period.
PRC represents such a phase shift $\Delta \phi$ as a function of the phase $\phi$ when the stimulus is applied.
We computed PRC by transiently changing the inverse temperature from ${\beta_1}$ to ${\beta_2}$ for one time unit (see Fig.\ref{fig1}C), by defining the phase of oscillation by the time when the total phosphorylation level takes maximum at $\phi = 0, 2 \pi, \cdots$.
As an indicator of the plasticity of phase, we measured the difference between maximum and minimum values of  the phase change $\Delta \phi$ in PRC \cite{foot2} normalized by the magnitude of a stimulus by fixing its duration as one time unit.
The dependence of $\Delta \phi$ and $\Delta T / T$ on $E_{dp}$ with fixed $E_p$ is plotted in Fig.\ref{fig1}A.
When $E_{dp}$ is low, i.e., when the temperature dependence of dephosphorylation reaction is weak, $\Delta T / T$ is small and $\Delta \phi$ is large.
This reciprocity was also obtained against changes in other parameters, $\beta_1$ and $A_{total}$ (see Fig.3 of \cite{Supply}).
This indicates that a biochemical oscillator with a homeostatic period against an environmental change can easily shift its phase under the same environmental change.
We also confirmed such reciprocity against change in ATP, i.e., the case of nutrient compensation (see Fig.5 of \cite{Supply}).

\begin{figure}[]
\includegraphics{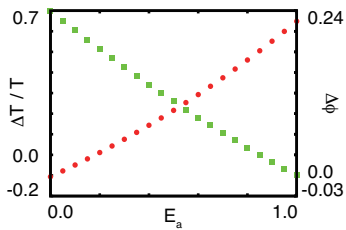}
\caption{
Reciprocity between the robustness of period and plasticity of phase in the TTO model.
Difference between periods at two temperatures ($\beta_1 = 0.0$ and $\beta_2 = 0.5$) ($\Delta T / T$, red circle) and the amplitude of the phase response curve against a transient jump of temperature from $\beta_1$ to $\beta_2$ ($\Delta \phi$, green square) plotted against various $E_a$, while activation energies for other reactions are fixed at $1.0$.
$\Delta T / T$ and $\Delta \phi$ are calculated similarly to how they are calculated in Fig.\ref{fig1}.
}
\label{fig2}
\end{figure}

Now, we examine if such reciprocity holds in the other class of circadian clocks, the TTO.
In the TTO model, a clock-related gene is first transcribed and translated, and later such a translated protein represses the expression of its own gene with a time-delay.
When the transcription rate decreases, the amount of such protein also decreases, which weakens the suppression of the clock-related gene expression.
Consequently, such genes are transcribed again, leading to the oscillation of the gene expression level.
As a typical example of the TTO model, we choose here a model of a circadian clock of a fruit fly \cite{Kurosawa2005} (see model equations and Fig.1B of \cite{Supply}).
By varying the activation energy for mRNA degradation, $E_a$, and fixing activation energies for other reactions, we measured $\Delta \phi$ and $\Delta T / T$ using the same procedure as in the Kai model.
Then, $\Delta T / T$ is low and $\Delta \phi$ is high for a low $E_a$ value, and $\Delta T / T$ ($\Delta \phi$) increases (decreases) with the increase in $E_a$ (Fig.\ref{fig2} and see also Fig.7 of \cite{Supply}).
Thus, the reciprocity holds also in the TTO.

To discuss the reciprocity analytically, we then study the Stuart--Landau model, a minimal model for simple sinusoidal oscillation \cite{Kuramoto1984}.
The model consists of the amplitude $R$ and argument $\Theta$, where $R$ and $\dot{\Theta}$ reach a constant value at the limit-cycle attractor.
Indeed, this model is derived as a normal form close to the Hopf bifurcation point.
We introduce an external parameter $\beta$:
\begin{subequations}
\begin{eqnarray}
\frac{dR(\beta)}{dt} &=& f_1(\beta) R - R^3, \label{drdt} \\
\frac{d\Theta(\beta)}{dt} &=& f_1(\beta) \omega + f_2(\beta) R^2, \label{dtdt}
\end{eqnarray}
\end{subequations}
where, $f_1(\beta)$ is a response function of the first order term in complex Ginzburg-Landau equation, and $f_2(\beta)$ is that of the third order term (for choice of each functions, see \cite{Supply}).
Considering the stability of limit cycle, the relaxation of $R$ after perturbation is assumed to be much faster than that of $\Theta$ \cite{foot3}.
Here, the period is given as:
\begin{equation}
T(\beta) = 2 \pi \left\{ f_1(\beta) \left( \omega + f_2(\beta) \right) \right\}^{-1}.
\end{equation}
Thus, after the change $\beta \rightarrow \beta + \Delta \beta$, the dependence of period on $\beta$ is given as:
\begin{equation}
\Delta \ln T(\beta) \simeq - \Delta \ln f_1(\beta) - \Delta f_2(\beta) \left( \omega + f_2(\beta) \right)^{-1} . \label{dt}
\end{equation}
Here, we neglected higher order terms of $\Delta \beta$, assuming that it is sufficiently smaller than $\beta$.
From Eq. (\ref{dt}), if $\Delta \ln f_1(\beta) = - \Delta f_2(\beta) \left( \omega + f_2(\beta) \right)^{-1}$ (i.e., $f_1' / f_1 = - f_2' / (\omega + f_2)$) is satisfied, the dependence of the period on $f_2(\beta)$ will be counterbalanced by $f_1(\beta)$, and the period is compensated against a change in $\beta$.

The argument $\Theta$ is defined only on a limit-cycle orbit, and we introduce the phase $\phi$ to extend the definition to the phase space out of the limit-cycle attractor, in particular to its basin.
It is postulated that $\phi$ agrees with $\Theta$ on the limit-cycle orbit, i.e., different orbits from the same $\phi$ converge to the same point on the limit cycle having the same $\Theta$ value.
Now, we will derive an isochrone, which is a set of points with the same $\phi$ on the phase space.
$\phi$ is expected to have rotational symmetry, hence the isochrone of the Stuart--Landau equation against the parameter $\beta$ is derived as
\begin{equation}
\phi(R, \Theta, \beta) = \Theta + f_2(\beta) \left\{ \ln R - \frac{1}{2} \ln f_1(\beta) \right\}. \label{phicomp}
\end{equation}
(see a supplemental text of \cite{Supply}.)
Then, we consider an operation that increases $\beta$ from $\beta_0$ to $\beta_0 + \Delta \beta$ and instantaneously reverses it to $\beta_0$.
By assuming that $R$  instantaneously relaxes to $R^*(\beta_0 + \Delta \beta) = (f_1(\beta_0 + \Delta \beta))^{1/2}$ while $\Theta$ remains unchanged, the phase after the above operation is derived as
\begin{equation}
\phi(\beta_0 + \Delta \beta) = \Theta(\beta_0) + \frac{f_2(\beta_0)}{2} \left\{\ln f_1(\beta_0 + \Delta \beta) - \ln f_1(\beta_0) \right\}.
\end{equation}
Hence, when $\Delta \beta \ll \beta$, the change in phase is derived as:
\begin{equation}
\Delta \phi(\beta_0) = f_2(\beta_0) \Delta \ln f_1(\beta) / 2. \label{dp}
\end{equation}
Therefore, from Eqs. (\ref{dt}) and (\ref{dp}), changes in the period and phase are represented by an equality.
\begin{equation}
a \Delta \ln T + \Delta \phi = c \label{rec},
\end{equation}
where $a = f_2(\beta) / 2$, $c = - f_2(\beta) \Delta f_2(\beta) / 2 \left( \omega + f_2(\beta) \right)$, which depend only on $f_2(\beta)$ and not on $f_1(\beta)$.
Thus, when we construct $f_1(\beta)$, which compensates for the dependence of $f_2(\beta)$ on $\beta$ according to Eq. (\ref{dt}), the phase is altered as $\Delta \phi = c$.
On the other hand, when $f_1(\beta)$ is independent of $\beta$, the phase is also independent due to Eq. (\ref{dp}), while the period is strongly dependent on $\beta$ as $\Delta \ln T = c / a = -\Delta f_2(\beta) / (\omega + f_2(\beta))$.

We also confirmed the reciprocity is valid in the modified van der Pol oscillator \cite{Vanderpol1926} with strong nonlinearlity, i.e., beyond the neighborhood of Hopf bifurcation (See Fig.9 of \cite{Supply}).

\begin{figure}[]
\includegraphics{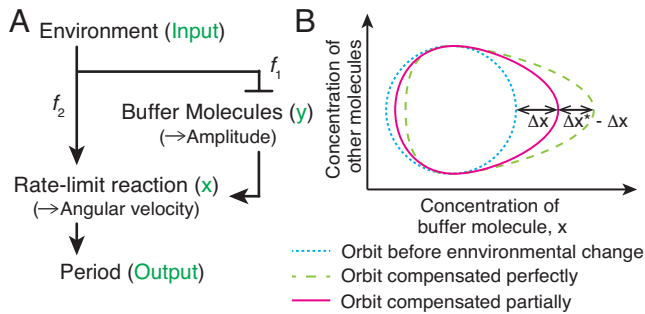}
\caption{
Schemes of the reciprocity between the robustness of period and plasticity of phase.
(A) Schematic networks of a generic (bio)chemical oscillator exhibiting homeostasis of period.
Pointed and flat arrowheads indicate positive and negative regulation, respectively.
Correspondences with a simple feedforward adaptation motif are represented by green characters in parentheses.
(B) Scheme of limit-cycle orbits with compensation of the period against environmental change.
Blue dotted line is a stable limit-cycle orbit before environmental change.
Green dashed line and magenta solid line are stable limit-cycle orbits after environmental change when the period is perfectly compensated and partially compensated, respectively.
}
\label{fig3}
\end{figure}

The origin of reciprocity is also understood from the viewpoint of adaptation motif.
The standard minimal feedforward motif for adaptation consists of two components, $x$ and $y$ \cite{Alon2006}.
In the feedforward network in Fig.\ref{fig3}A, an input changes both components $x$ and $y$, while $y$ gives an input to $x$.
Here, the direct path to $x$ and the indirect path via $y$ from the input have opposite signs.
Then, the response of the output $x$ via the direct path is later canceled by $y$, and the adaptation behavior against the input is shaped.
The degree of adaptation depends on the strength of the indirect regulation; weak regulation induces a partial adaptation and strong regulation leading to the cancellation of the two paths, induces perfect adaptation \cite{Koshland1982, Segel1986}.

Our Stuart--Landau model also has a feedforward motif consisting of amplitude and angular velocity.
When an environmental condition $\beta$ is changed, the angular velocity and amplitude are altered by the terms $f_1 (\beta)$ and $f_2 (\beta)$.
After a direct change in angular velocity, the change is relaxed by the change in amplitude.
The period is determined as the inverse of the angular velocity.
If changes in the amplitude are large, period is perfectly compensated and phase is plastic.
In contrast, if the change in amplitude is small, the angular velocity shows partial adaptation leading to partial compensation of the period while the phase is only slightly altered.
Therefore, the reciprocity is understood as the adaptation dynamics on a limit cycle.

Indeed, the above argument of the adaptation on the limit cycle generally holds, for PTO and TTO models, where we can generally consider the scheme of Fig.\ref{fig3}A.
Environmental change directly influences the angular velocity while it is also buffered in the amplitude and then influences the phase.
In a biochemical clock, the period mainly depends on the rate-limit reactions, which are slower than others.
Environmental change will alter the speed of such rate-limit reactions, which is later counterbalanced by the change in the concentration of buffer molecules \cite{foot4}.
In fact, in the PTO model, the amount of free enzyme working as a buffer molecule can counterbalance the speed of the rate-limit reaction.
Hence, the period of the oscillator is homeostatic against environmental changes.
Likewise, in TTO model, mRNA plays the role of such buffer molecule.

In this time, the limit-cycle orbit of oscillators with compensation shifts in the phase space of chemical concentrations to change that of a buffer molecule (see Fig.\ref{fig3}B).
When homeostatic response is achieved, the concentration of a buffer molecule $x$ should be changed with $\Delta x$ by the change in the external environment.
Then the limit-cycle orbit will be shifted to change the concentration of a buffer molecule, and the magnitude of such shift and the change in isocline will be $O (\Delta x)$ considering that continuous change in the isocline against $\Delta x$ which is small.
Then, $\Delta \phi \propto \Delta x$ is expected.
On the other hand, when the change in the concentration of a buffer molecule is not sufficient to counterbalance the environmental stimulus, the concentrations of other molecules will change.
Let us represent the concentration of $x$ needed for perfect adaptation as $\Delta x^*$.
Then, the change in the concentration of the other molecule of the lowest order is proportional to $\Delta x^* - \Delta x$.
The period also changes accordingly, so that $\Delta T / T \propto \Delta x^* - \Delta x$ is expected.
By combining the two proportionally relationships, we obtain $a \Delta \phi + b \Delta T / T = \Delta x^*$ with coefficients of proportionality $a$ and $b$.

We have shown that reciprocity exists in both the PTO and TTO models.
The currently known mechanisms of circadian oscillation can be classified into the above two cases \cite{Qin2010}, and the reciprocity is expected to be achieved universally in circadian clocks \cite{foot6}.
In a circadian clock system of a mold, {\it Neurospora crassa}, it was reported that a loss-of-temperature-compensation mutant, {\it frq-7}, shows smaller phase shift against transient temperature change than the wild type \cite{Lakin-Thomas1990, Nakashima1987, Rensing1987a}.
Although the quantitative relationship between temperature compensation and phase plasticity was not investigated therein, we expect that a quantitative experiment will confirm our reciprocity, not only in {\it Neurospora crassa} but also in other organisms in which loss-of-temperature-compensation mutants are isolated, e.g., fruit fly \cite{Matsumoto1999} and cyanobacteria \cite{Murayama2010}.
Here, we demonstrated the reciprocity against changes in the temperature and the nutrient concentration, but from theoretical consideration, it is expected to hold generally against a variety of stimuli, such as the change in strength of light and transcription rate \cite{Dibner2009, Kim2012}, as long as the adaptation mechanism works.
Moreover, it is also expected that the reciprocity is not limited to the circadian clock; it holds generally as long as the adaptation mechanism with buffering molecules works {\cite{foot7}.
Our reciprocity will give a general quantitative law for such adaptation systems.

\begin{acknowledgements}
This work was partially supported by the Platform for Dynamic Approaches to Living System from MEXT, Japan; Dynamical Micro-scale Reaction Environment Project, JST; and JSPS KAKENHI Grant No. 15K18512.
The authors would like to thank B. Pfeuty, H. Kori, K. Fujimoto, and U. Alon for useful discussion.
\end{acknowledgements}

\newpage
\setcounter{figure}{0}
\setcounter{equation}{0}

\section{Supplemental material}

\section{Models}
\subsection{Post-translational oscillator (PTO) model}
We introduce the KaiC allosteric model \cite{VanZon2007} (Fig. \ref{figs1}A).
The KaiC protein has six monomers, and each monomer has multiple phosphorylation sites.
Here, we assumed each KaiC monomers have only two phosphorylation states, phosphorylated and unphosphorylated.
KaiC hexamer takes an active or inactive form.
By denoting $C_i$ and $\tilde{C}_i$ as active and inactive forms with the $i$ phosphorylated monomers, respectively, their temporal changes are given as:

\begin{subequations}
\begin{eqnarray}
\frac{d[C_i]}{dt} &=& (1 - \delta_{i,0}) \frac{k_p [A] [C_{i-1}]}{K_{i-1} + [A]} - (1 - \delta_{i,6}) \frac{k_p [A] [C_i]}{K_i + [A]} \nonumber \\
& & + \delta_{i,0} b [\tilde{C}_i] - \delta_{i,6} f [C_i], \\
\frac{d[\tilde{C}_i]}{dt} &=& k_{dp} ((1 - \delta_{i,6}) [\tilde{C}_{i+1}] - (1 - \delta_{i,0}) [\tilde{C}_i]) \nonumber \\
& & - \delta_{i,0} b [\tilde{C}_i] + \delta_{i,6} f [C_i], \\
A_{total} &=& [A] + \sum_{i=0}^{5} \frac{[A] [C_i]}{K_i + [A]},
\end{eqnarray}
\end{subequations}
where $A$ denotes the free KaiA protein that works as an enzyme for phosphorylation.
$A_{total}$ is the total KaiA amount, which is a constant because the total amounts of both KaiC and KaiA are conserved quantities, and $[x]$ denotes the concentration of $x$.
$K_i$ is the dissociation constant between $C_i$ and $A$.
$k_p$ and $k_{dp}$ denote the rate of phosphorylation and dephosphorylation of KaiC, respectively, which depend on temperature as $k_p \propto \exp(- \beta E_p)$ and $k_{dp} \propto \exp(- \beta E_{dp})$, where $E_p$ ($E_{dp}$) is the activation energy for phosphorylation (dephosphorylatiion) and $\beta$ is the inverse temperature by taking the Boltzmann constant as unity.
For case of the nutrient compensation (Fig. \ref{figs5}), we considered the model where only the phosphorylation reaction speed, $k_p$, is proportional to ATP-to-ADP ratio and others are independent of it.

\begin{figure}[]
\includegraphics{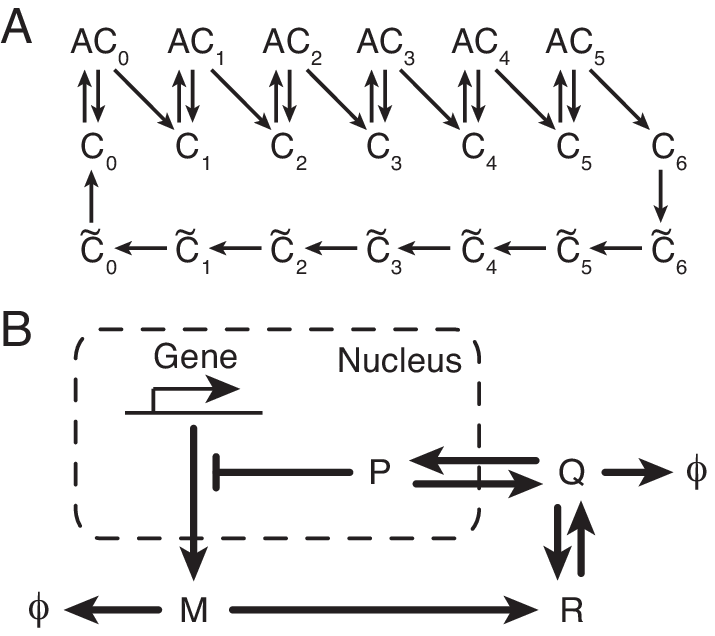}
\caption{
Schemes of models of the circadian clock.
(A) KaiC allosteric model as an example of the PTO model.
$C_i$ and $\tilde{C_i}$ are an active and inactive substrate (KaiC) with $i$ phosphorylated residues, respectively.
$A$ is an enzyme (KaiA) and $AC_i$ is an enzyme-substrate complex with $i$ phosphorylated residues.
(B) TTO model.
$M$ is an mRNA and $R$ is the precursor of a protein.
$Q$ and $P$ are an extranuclear and internuclear protein, respectively.
$P$ can negatively regulate gene expression and create a negative feedback loop.
}
\label{figs1}
\end{figure}

\subsection{Transcription-translation-based oscillator (TTO) model}\
We introduce a model of a circadian clock of a fruit fly \citep{Kurosawa2005} (Fig.\ref{figs1}B).
The governing differential equations are described below:
\begin{subequations}
\begin{eqnarray}
\frac{d[M]}{dt} &=& \frac{k}{h+[P]} - \frac{a[M]}{a'+[M]}, \\
\frac{d[R]}{dt} &=& \frac{s[M]}{s'+[M]} -\frac{b[R]}{b'+[R]} +\frac{c[Q]}{c'+[Q]}, \\
\frac{d[Q]}{dt} &=& \frac{b[R]}{b'+[R]} -\frac{c[Q]}{c'+[Q]} -\frac{d[Q]}{d'+[Q]} \\
& & -\frac{u[Q]}{u'+[Q]} +\frac{v[P]}{v'+[P]}, \nonumber \\
\frac{d[P]}{dt} &=& \frac{u[Q]}{u'+[Q]} -\frac{v[P]}{v'+[P]},
\end{eqnarray}
\end{subequations}
where $M$ is the mRNA of the clock-related gene ({\it per} mRNA); $R$ is the protein precursor of a clock-related protein (PER protein); $Q$ and $P$ are an extranuclear protein and a nucleic protein, respectively; and $[x]$ denotes the concentration of $x$.
$k$, $a$, $s$, $b$, $c$, $d$, $u$, and $v$ are rate constants, and $h$, $a'$, $s'$, $b'$, $c'$, $d'$, $u'$, and $v'$ are dissociation constants.

In \citep{Kurosawa2005}, it was reported that $a$ and $k$ are especially important to determine the length of the period.
Following this report, we set the rate constant of each reaction to follow the Arrhenius equation, i.e., a kinetic constant of mRNA degradation as $a \propto \exp(-\beta E_a)$ and that of transcription as $k \propto \exp(-\beta E_k)$ where $E_a$ and $E_k$ are activation energies of mRNA degradation and transcription, respectively.

\begin{figure}[]
\includegraphics[clip]{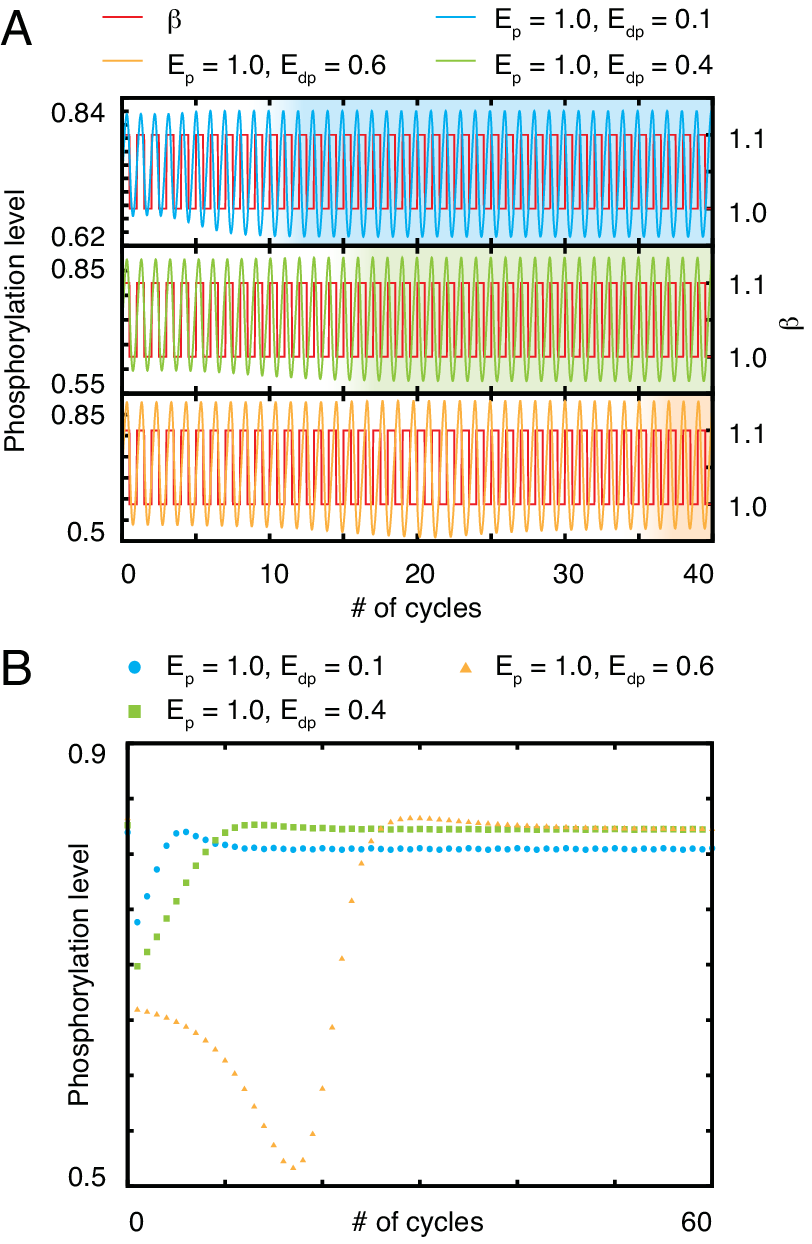}
\caption{
Entrainment of the clock in the PTO model against external temperature cycle,
For different values of activation energy of dephosphorylation reactions $E_{dp}$.
(A) Time evolution of phosphorylation level ($\Sigma i [C_i] / 6 C_{total}$) with $E_{dp} =$ 0.1 (cyan line), 0.4 (green line), and 0.6 (orange line), plotted against the time, normalized by the period of temperature cycles.
As shown with the red line, the temperature (to be precise, its inverse $\beta$) is periodically changed between $\beta_1 = 1.0$ and $\beta_2 = 1.1$ with the period at $\beta_1$, where each interval at $\beta_1$ and $\beta_2$ is set identical.
The phosphorylation oscillation is entrained with the external temperature cycle, at the time shaded in the figure.
 (B) Plots of phosphorylation levels per period of the external cycle, i.e., at the time when the temperature is switched from $\beta_1$ to $\beta_2$.
Initially, the phosphorylation level changes per period, and then, after entrainment, phosphorylation level takes an almost constant, when the clock is entrained with the external cycle.
For smaller $E_{dp}$, the phosphorylation clock is entrained faster.
}
\label{figs2}
\end{figure}

\begin{figure}[]
\includegraphics[clip]{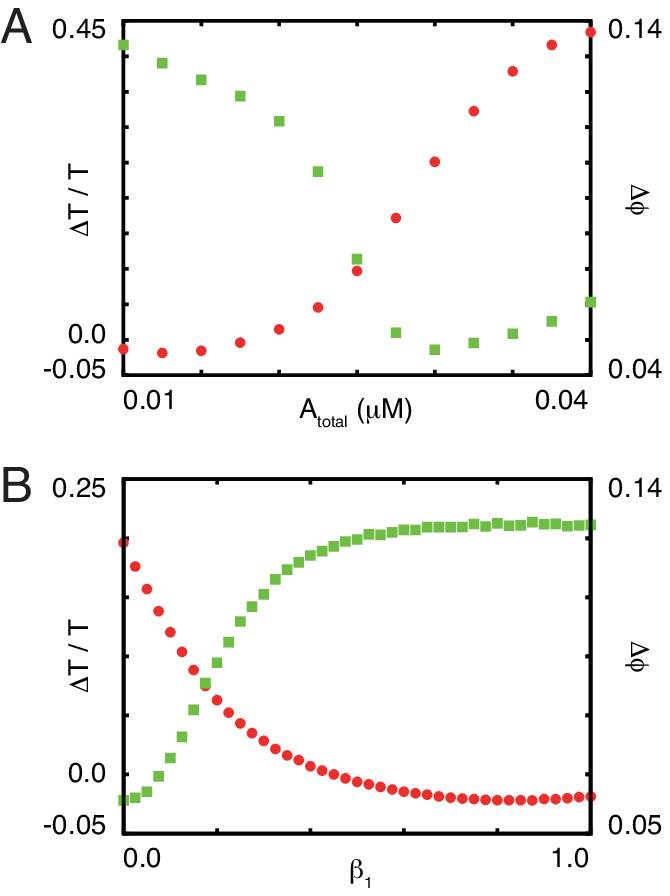}
\caption{
Reciprocity between the robustness of period and plasticity of phase in the PTO model under various enzyme concentrations and under various temperatures.
(A) Difference between periods at two temperatures ($\beta_1 = 1.0$ and $\beta_2 = 1.5$) ($\Delta T / T$, red circle) and the amplitude of the phase response curve against a transient jump of temperature from $\beta_1$ to $\beta_2$ ($\Delta \phi$, green square) plotted against various total concentrations of the enzyme.
(B) Difference between periods at two temperatures ($\beta_1$ and $\beta_2 = \beta_1 + 0.5$) ($\Delta T / T$, red circle) and the amplitude of the phase response curve against a transient jump of temperature from $\beta_1$ to $\beta_2$ ($\Delta \phi$, green square) plotted against various $\beta_1$.
$\Delta T / T$ and $\Delta \phi$ are calculated similarly to how they are calculated in Fig.1 in the main text.
}
\label{figs3}
\end{figure}

\begin{figure}[]
\includegraphics[clip]{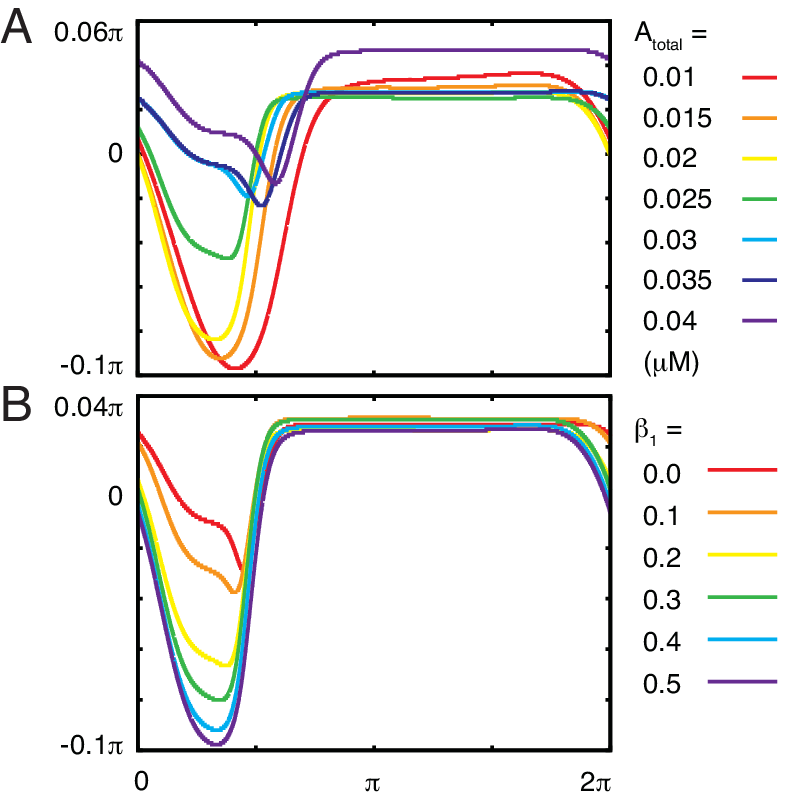}
\caption{Phase response curve of the PTO model against the transient increase in $\beta$ for different values of (A) $A_{total}$ and (B) $\beta_1$, the inverse of temperature.
PRCs are calculated in the same manner as in Fig.1C in the main text.
(A) Lines of different colors represent PRCs for different values of the total KaiA $A_{total}$; $A_{total} = $ 0.01 (red line), 0.015 (orange line), 0.02 (yellow line), 0.025 (green line), 0.03 (cyan line), 0.035 (indigo line), and 0.04 $\mathrm{\mu M}$ (purple line).
(B) Lines of different colors represent PRCs for different values of $\beta_1$, while the inverse temperature is changed transiently to $\beta_2 = \beta_1 + 0.5$; $\beta_1 = $ 0.0 (red line), 0.1 (orange line), 0.2 (yellow line), 0.3 (green line), 0.4 (cyan line), and 0.5 (purple line).
}
\label{figs4}
\end{figure}

\begin{figure}[]
\includegraphics[clip]{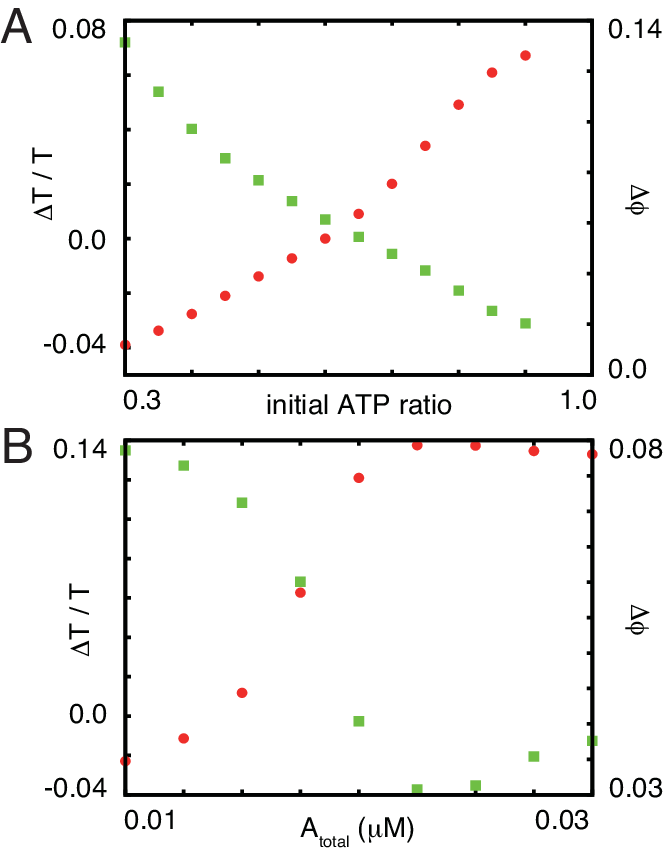}
\caption{
Reciprocity between the robustness of period and plasticity of phase in the PTO model against the change in ATP ratio ($[{\rm ATP}] / ([{\rm ATP}] + [{\rm ADP}])$) and enzyme concentrations.
(A) Difference between periods ($\Delta T / T$, red circle) for two ATP ratios (initial ATP ratio $x$ and $x + 0.1$), and the amplitude of the phase response curve against a transient jump of ATP ratio from $x$ to $x + 0.1$ ($\Delta \phi$, green square) plotted against different values of initial ATP ratio $x$.
(B) Difference between periods ($\Delta T / T$, red circle) for two ATP ratios ($x$ and $x + 0.1$), and the amplitude of the phase response curve against a transient jump of ATP ratio from $x$ to $x + 0.1$ ($\Delta \phi$, green square) plotted against different values of total concentrations of the enzyme.
Here, we set initial ATP ratio $x$ as 0.5.
$\Delta T / T$ and $\Delta \phi$ are calculated similarly to how they are calculated in Fig.1 in the main text.
}
\label{figs5}
\end{figure}

\begin{figure}[]
\includegraphics[clip]{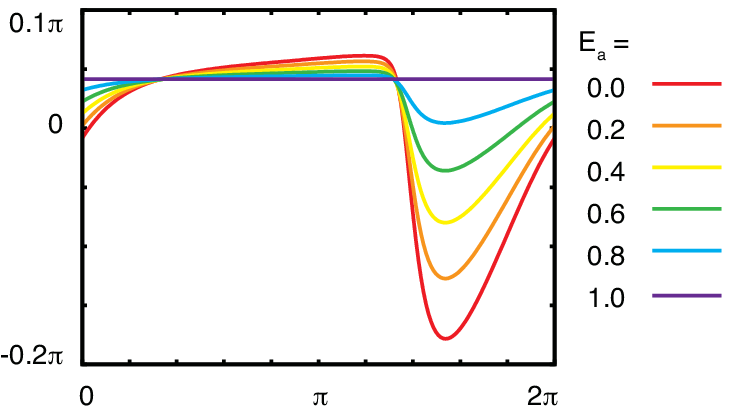}
\caption{Phase response curve of transcription-translational-based oscillator (TTO) model against the transient increase in $\beta$ for different values of $E_{a}$.
The zero-phase point $\phi = 0, 2 \pi$ is defined as the state in which the total amount of protein ($[R] + [Q] + [P]$) takes its maximal value, and the phase increases from 0 to $2 \pi$ proportionally to time.
The inverse temperature $\beta$ is increased from $\beta_1 = 0.0$ to $\beta_2 = \beta_1 + 0.5 = 0.5$ for the duration of one unit of time.
Lines of different colors represent PRCs for different activation energies for mRNA degradation $E_{a}$: $E_{a} = $ 0.0 (red line), 0.2 (orange line), 0.4 (yellow line), 0.6 (green line), 0.8 (cyan line), and 1.0 (purple line).
}
\label{figs6}
\end{figure}

\begin{figure}[]
\includegraphics[clip]{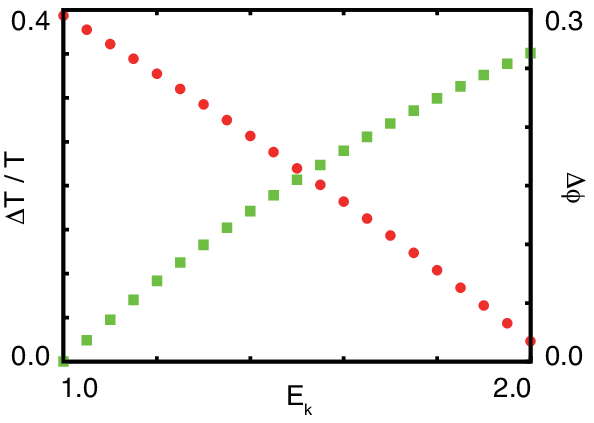}
\caption{
Reciprocity between the robustness of period and plasticity of phase in the TTO model.
Difference between periods at two temperatures ($\beta_1 = 0.0$ and $\beta_2 = 0.5$) ($\Delta T / T$, red circle) and the amplitude of the phase response curve against a transient jump of temperature from $\beta_1$ to $\beta_2$ ($\Delta \phi$, green square) plotted against various $E_k$ where activation energies for other reactions are fixed at $1.0$.
$\Delta T / T$ and $\Delta \phi$ are calculated similarly to how they are calculated in Fig.1 in the main text.
}
\label{figs7}
\end{figure}

\section{Calculation of the entrainability}
To calculate the entrainability in Fig.1B in the main text, initially, 36 oscillators are set at same intervals of the phase, and the number of cycles needed for the oscillators to synchronize is computed.
As for the numerical criteria, and we regard that the clock is entrained if the circular variance of phase from different initial conditions is smaller than $10^{-9}$.
Here, the temperature cycle is applied as a square wave between $\beta_1 = 1.0$ and $\beta_2 = 1.1$ with an equal interval, with the period at the condition of $\beta = \beta_1$. 
When entrainability is zero, the oscillator is never entrained.
See also Fig.2.

\section{Analysis of Stuart-Landau Equation}
We introduce the Stuart-Landau equation with an external parameter $\beta$:
\begin{subequations}
\begin{eqnarray}
\frac{dR(\beta)}{dt} &=& f_1(\beta) R - R^3, \label{drdt} \\
\frac{d\Theta(\beta)}{dt} &=& f_1(\beta) \omega + f_2(\beta) R^2. \label{dtdt}
\end{eqnarray}
\end{subequations}
This form is derived from the complex Ginzburg-Landau equation, where $f_1(\beta)$ represents the change in the bifurcation parameter, and $f_2(\beta)$ in that for phase-amplitude coupling.
\begin{equation}
\frac{dA}{dt} = f_1(\beta) (1 + i \omega) A - (1 - i  f_2(\beta)) |A|^2 A ,
\end{equation}
where, the first order term ($f_1(\beta)$) and the third order term ($f_2(\beta)$) should have different dependency upon $\beta$ for the adaptation mechanism to work.
Indeed the reciprocity holds generally for other forms of the third order term to satisfy adaptation.

Considering the stability of limit cycle, the relaxation of $R$ after perturbation is assumed to be much faster than that of $\Theta$.
Hence, $R^*$, the steady-state value of $R$ ($>0$), is obtained as
\begin{equation}
R^{*}(\beta) = (f_1(\beta))^{1/2}. \label{rast}
\end{equation}
Here, the period is given as:
\begin{equation}
T(\beta) = 2 \pi \left\{ f_1(\beta) \left( \omega + f_2(\beta) \right) \right\}^{-1}.
\end{equation}

Here, the argument $\Theta$ is defined only on a limit-cycle orbit, and we introduce the phase $\phi$ to extend the definition to the phase space out of the limit-cycle attractor, in particular to its basin.
It is postulated that $\phi$ agrees with $\Theta$ on the limit-cycle orbit, i.e., different orbits from the same $\phi$ converge to the same point on the limit cycle having the same $\Theta$ value.
Now, we will derive an isochrone, which is a set of points with the same $\phi$ on the phase space.
$\phi$ is expected to have rotational symmetry and is given as
\begin{equation}
\phi(R, \Theta, \beta) = \Theta(\beta) + g(R(\beta)). \label{phi} 
\end{equation}
Moreover, the time evolution of $\phi$ should coincide with that of $\Theta$.
Thus, 
\begin{equation}
\frac{d \phi}{dt} = f_1(\beta) \omega + f_1(\beta) f_2(\beta) . \label{dphidt}
\end{equation}
From Eqs. (\ref{drdt}), (\ref{dtdt}), (\ref{phi}), and (\ref{dphidt}), the time evolution of $\phi$ is derived as:
\begin{eqnarray}
\frac{d \phi(R, \Theta, \beta)}{dt} &=& \frac{d \Theta}{dt} + \frac{dR}{dt} \frac{d g(R)}{dR} \nonumber \\
&=& f_1(\beta) \omega + f_2(\beta) R^2 \nonumber \\
& & + \{ f_1(\beta) R - R^3 \} \frac{dg(R)}{dR} .
\end{eqnarray}
Hence, 
\begin{equation}
\{ f_1(\beta) R - R^3 \} \frac{dg(R)}{dR} = f_2(\beta) \{ f_1(\beta) - R^2 \}.
\end{equation}
Thus,
\begin{equation}
\frac{dg(R)}{dR} = \frac{f_2(\beta)}{R}.
\end{equation}
Accordingly, we obtain
\begin{equation}
g(R) = f_2(\beta) \ln R + C \label{gr}.
\end{equation}
Next, we will derive $C$.
Because the phase $\phi(R, \Theta, \beta)$ should coincide with $\Theta$ at $R = R^*$, $g(R^*)$ becomes zero, and then from Eq.(4) in the main text and Eq.(\ref{gr}), 
\begin{eqnarray}
C &=& - \frac{1}{2} f_2(\beta) \ln f_1(\beta), \nonumber \\
\phi(R, \Theta, \beta) &=& \Theta + f_2(\beta) \left\{ \ln R - \frac{1}{2} \ln f_1(\beta) \right\}. \label{phicomp}
\end{eqnarray}
Thus, the isochrone of the Stuart--Landau equation against the parameter $\beta$ is derived.
Then, we consider an operation that increases $\beta$ from $\beta_0$ to $\beta_0 + \Delta \beta$ and instantaneously reverses it to $\beta_0$.
At this time, by assuming that $R$  instantaneously relaxes to $R^*(\beta_0 + \Delta \beta) = (f_1(\beta_0 + \Delta \beta))^{1/2}$ while $\Theta$ remains unchanged, the phase after the above operation is derived as
\begin{equation}
\phi(\beta_0 + \Delta \beta) = \Theta(\beta_0) + \frac{f_2(\beta_0)}{2} \left\{\ln f_1(\beta_0 + \Delta \beta) - \ln f_1(\beta_0) \right\}.
\end{equation}
In contrast, the phase before the operation is given as
\begin{eqnarray}
\phi(\beta_0) = \Theta(\beta_0).
\end{eqnarray}
Hence, when $\Delta \beta \ll \beta$, the change in phase is derived as:
\begin{eqnarray}
\Delta \phi(\beta_0) &=& \phi(\beta_0 + \Delta \beta) - \phi(\beta_0) \nonumber \\
&=& f_2(\beta_0) \left\{ \ln f_1(\beta_0 + \Delta \beta) - \ln f_1(\beta_0) \right\} / 2 \nonumber \\
&=& f_2(\beta_0) \Delta \ln f_1(\beta) / 2. \label{dp}
\end{eqnarray}
Therefore, from Eq.(6) in the main text and Eq.(\ref{dp}), changes in the period and phase are represented by an equality.
\begin{equation}
a \Delta \ln T + \Delta \phi = c \label{rec},
\end{equation}
where $a = f_2(\beta) / 2$, $c = - f_2(\beta) \Delta f_2(\beta) / 2 \left( \omega + f_2(\beta) \right)$.
The reciprocity is generally true also for some different forms, such as:
\begin{subequations}
\begin{eqnarray}
\frac{dR(\beta)}{dt} &=& f_1(\beta) R - f_2(\beta) R^3, \label{exp1} \\
\frac{d\Theta(\beta)}{dt} &=& f_1(\beta) \omega + f_2(\beta) R^2,
\end{eqnarray}
\end{subequations}
and, 
\begin{subequations}
\begin{eqnarray}
\frac{dR(\beta)}{dt} &=& f_1(\beta) R - f_2(\beta) R^3, \label{exp2} \\
\frac{d\Theta(\beta)}{dt} &=& f_1(\beta) \omega + R^2.
\end{eqnarray}
\end{subequations}
In both the cases, Eq.(\ref{rec}) still holds, while the expression of $a$ and $c$ are different.
For Eq.(18), $a = 1 / 2$ and $c = - \Delta \ln f_2(\beta) / 2$.
For Eq.(19), $a = 1 / 2 f_2(\beta)$ and $c = \{ \Delta f_2(\beta) (f_2(\beta) \omega + 1)^{-1} - \Delta \ln f_2(\beta) \} / 2 f_2(\beta)$.

\section{van der Pol Oscillator Model with Parameter $\beta$}

\begin{figure}[]
\includegraphics[clip]{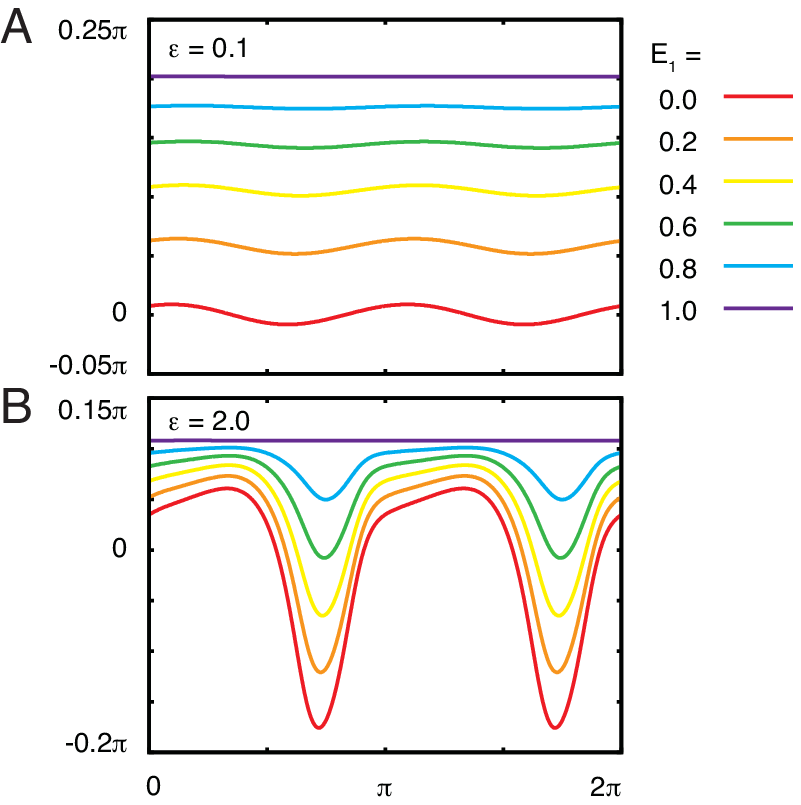}
\caption{Phase response curve of the van der Pol oscillator against the transient increase in $\beta$ for different values of $E_{1}$.
The zero-phase point $\phi = 0, 2 \pi$ is defined as the state in which $x$ takes its maximal value, and the phase increases from 0 to $2 \pi$ proportionally to time.
The inverse temperature $\beta$ increases from $\beta_1 = 0.0$ to $\beta_2 = \beta_1 + 0.5 = 0.5$ for the duration of one unit of time. 
The strength of nonlinearity $\epsilon$ is $\epsilon = $ 0.1 for (A) and  $\epsilon = $ 2.0 for (B).
Lines of different colors represent PRCs for different values of $E_{1}$: $E_{1} = $ 0.0 (red line), 0.2 (orange line), 0.4 (yellow line), 0.6 (green line), 0.8 (cyan line), and 1.0 (purple line).
}
\label{figs8}
\end{figure}

\begin{figure}[]
\includegraphics[clip]{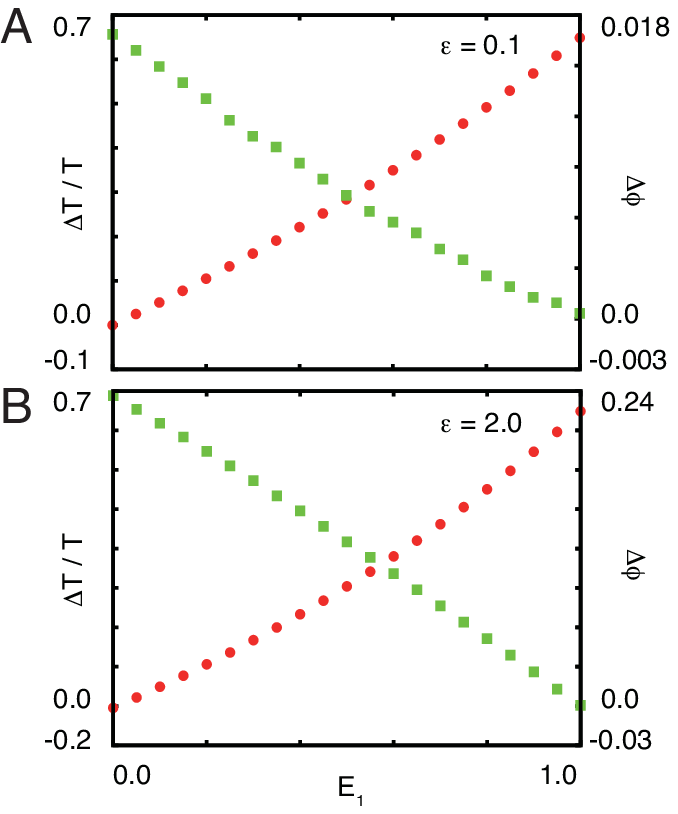}
\caption{
Reciprocity between the robustness of period and plasticity of phase in the van der Pol model.
Difference between periods at two temperatures ($\beta_1 = 0.0$ and $\beta_2 = 0.5$) ($\Delta T / T$, red circle) and the amplitude of the phase response curve against a transient jump of temperature from $\beta_1$ to $\beta_2$ ($\Delta \phi$, green square) are plotted against various $E_1$ with fixed $E_2 = 1.0$ in the van der Pol model with (A) $\epsilon = 0.1$ and (B) $\epsilon = 2.0$.
$\Delta T / T$ and $\Delta \phi$ are calculated in a similar way to Fig.1 in the main text.
}
\label{figs9}
\end{figure}

The van der Pol oscillator is one of the simplest nonlinear-oscillator models, and it is given by:
\begin{equation}
\frac{d^2x}{dt^2} - \epsilon (1  - x^{2}) \frac{dx}{dt} - b x = 0.
\end{equation}
The above equation can be decomposed into two ordinary differential equations as follows:
\begin{subequations}
\begin{eqnarray}
\frac{dx}{dt} &=& y, \\
\frac{dy}{dt} &=& \epsilon (1 - x^{2}) y - bx,
\end{eqnarray}
\end{subequations}
where $\epsilon$ is the strength of nonlinearity.
As $\epsilon$ increases, the system deviates more from a harmonic oscillator.

Here, we modify the van der Pol oscillator to show the change in the amplitude against a change in an external parameter as in the Stuart--Landau equation.
We alter van der Pol oscillator as
\begin{subequations}
\begin{eqnarray}
\frac{dx}{dt} &=& y, \\
\frac{dy}{dt} &=& \epsilon (f_1(\beta) - x^{2}) y - bx,
\end{eqnarray}
\end{subequations}
where $\beta$ is an environmental parameter.
If $\epsilon$ is sufficiently small, the amplitude $A$ can be derived by using perturbation calculation as
\begin{eqnarray}
\frac{dA}{dt} &=& \epsilon \left( \frac{f_1(\beta)}{2} A - \frac{|A|^2 A}{8} \right), \\
\left(|A|^*\right)^2 &=& f_1(\beta),
\end{eqnarray}
where $|A|^*$ is a fixed point value of $|A|$.
On the other hand, we introduce the dependence of the velocity on the environmental parameter as
\begin{subequations}
\begin{eqnarray}
\frac{1}{f_2(\beta)} \frac{dx}{dt} &=& y, \\
\frac{1}{f_2(\beta)} \frac{dy}{dt} &=& \epsilon (f_1(\beta) - x^{2}) y - bx.
\end{eqnarray}
\end{subequations}
When $\epsilon$ is small, the above modified van der Pol oscillator is expected to demonstrate same behavior as the Stuart--Landau model, as described in the main text.
When $\epsilon$ is large, however, the nonlinearity becomes large and the oscillatory behavior is altered from sinusoidal to relaxation.

To simulate the above model, we choose $f_1(\beta)$ and $f_2(\beta)$ as exponential forms similar to the Arrhenius equation in biochemical oscillators, i.e., $f_1(\beta) = e^{-\beta \Delta}$ and $f_2{\beta} = e^{-\beta E_2}$.
Then, the above equations are given as
\begin{subequations}
\begin{eqnarray}
\frac{dx}{dt} &=& e^{- \beta E_2} y, \\
\frac{dy}{dt} &=& \epsilon (e^{- \beta E_1} - e^{- \beta E_2} x^{2}) y - e^{- \beta E_2} b x,
\end{eqnarray}
\end{subequations}
where $E_1$ is given by $E_1 = \Delta + E_2$.
We use the above equations.

When the intensity of nonlinearity, $\epsilon$, is small, the magnitude of change in the period and magnitude of change in the phase are fitted well by a linear relationship $a \Delta T / T + b \Delta \phi = c$ ($a$, $b$, and $c$ are constants) (see Fig.\ref{figs9}A).
Here, as the intensity of nonlinearity increases, the orbit of a limit cycle is deformed from a circle, and the dynamics shifts from sinusoidal oscillation to relaxation oscillation.
Still, the reciprocity is valid as long as the magnitude of stimuli is not exceedingly large (see Fig.\ref{figs9}B).
Thus, the reciprocity is a universal feature beyond the neighborhood of Hopf bifurcation.


\begin{thebibliography}{99}
\bibitem{foot2}
$\Delta \phi$ is thought to be less model-dependent than the other measures including the entrainability and the shape of PRC.
\bibitem{foot3}
If the system is in the vicinity of Hopf bifurcation, the relaxation of $R$ may be slowed down, and further study is necessary.
\bibitem{foot7}
In weakly coupled oscillators under small perturbation or in the vicinity of Hopf bifurcation, robustness of period can be achieved without the adaptation mechanism.
Such oscillators without the adaptation mechanisms do not (necessarily) satisfy the reciprocity.
\end{thebibliography}
\end{document}